\documentclass[fleqn,twoside]{article}
\usepackage{espcrc2}

\usepackage{graphicx}
\usepackage[figuresright]{rotating}
\usepackage{amsmath,amssymb,epsf,epsfig,a4wide}

\def\gappeq{\mathrel{\rlap {\raise.5ex\hbox{$>$}}
{\lower.5ex\hbox{$\sim$}}}}
\def\lappeq{\mathrel{\rlap{\raise.5ex\hbox{$<$}}
{\lower.5ex\hbox{$\sim$}}}}

\title{High-Energy Astrophysics and Cosmology}

\author{John~Ellis\address[CERN]{Theory Division, CERN, \\
CH 1211 Geneva 23, Switzerland}%
        \thanks{Talk presented at the XIIth International Symposium on 
Very-High-Energy Cosmic-Ray Interactions, CERN, July 2002}}
       
\begin{document}

\begin{abstract}
Interfaces between high-energy physics, astrophysics and cosmology are 
reviewed, with particular emphasis on the important roles played by 
high-energy cosmic-ray physics. These include the understanding of 
atmospheric neutrinos, the search for massive cold dark matter particles 
and possible tests of models of quantum gravity. In return, experiments at 
the LHC may be useful for refining models of ultra-high-energy cosmic 
rays, and thereby contributing indirectly to understanding their origin. 
Only future experiments will be able to tell whether these are due to some 
bottom-up astrophysical mechanism or some top-down cosmological mechanism.
\vspace{1pc}
\begin{center}
CERN-TH/2002-300 ~~~~~~~~~~~~~~~ astro-ph/0210580
\end{center}
\end{abstract}

\maketitle

\section{Introduction}

\begin{figure}[htb]
\centering
\includegraphics*[width=85mm]{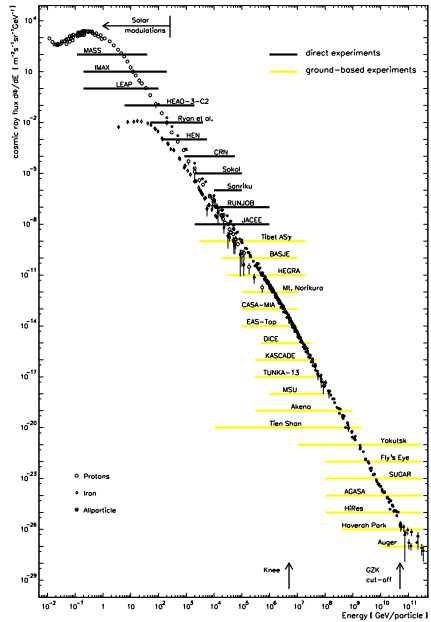}
\vspace{-2cm}
\caption{
The cosmic-ray spectrum extends over many decades in energy, which are 
sampled by many different experiments~\cite{WS}.}
\label{fig:CR}
\end{figure}

There are two ways in which high-energy particles have appeared naturally 
in the Universe: one is via energetic astrophysical sources such as 
gamma-ray bursters (GRBs) or active galactic nuclei (AGNs), and the other 
is via the characteristically high particle energies in the very early 
Universe. In this talk, I illustrate the possible r\^oles of both these 
types of sources, and discuss some related open questions in relation to 
the cosmic-ray energy spectrum shown in Fig.~\ref{fig:CR}. Here are a few 
examples.

Even relatively `low-energy' parts of this spectrum in the range $1 \sim
10$~GeV are directly connected to ultra-high-energy particle physics, via
their r\^ole in producing atmospheric neutrinos~\cite{SK}, one of our
windows on Grand Unified Theories (GUTs). Somewhat higher-energy parts of
the spectrum up to energies $\sim 1$~TeV are relevant to the experimental
search for cold dark matter particles. Confirmed observations of GRBs have
been limited to the MeV energy range, but there are unconfirmed reports of
observations in the GeV and even TeV~\cite{LANL} energy
ranges, and GRBs might even be responsible
for the ultra-high-energy cosmic rays (UHECRs). Alternatively, these might 
be due to some
exotic top-down mechanism involving the decays of supermassive particles
produced in the very early Universe~\cite{cryptons}, or some other
extreme astrophysical sources. Either way, they may provide a unique
laboratory to look for violations of fundamental principles such as
Lorentz invariance~\cite{GM}.

Before discussing these possible playgrounds for high-energy physics, let 
us first review the basis for the Big-Bang cosmology that plays an 
essential r\^ole in the following sections of this talk.

\section{Big-Bang Cosmology}

According to standard Big-Bang cosmology, the entire visible Universe is
expanding homogeneously and isotropically from a very dense and hot
initial state. Apart from the present Hubble expansion, the first piece of
evidence for the Big Bang was the cosmic microwave background (CMB)
radiation, which is thought to have been emitted when the Universe was
about 3000 times smaller and hotter than it is today, with age $\sim 3
\times 10^5$~y. The CMB has a dipole deviation from isotropy at the
$10^{-3}$ level, which this is believed to be due to the Earth's motion
relative to a Machian cosmological frame.

\begin{figure}[htb]
\centering
\includegraphics*[width=65mm]{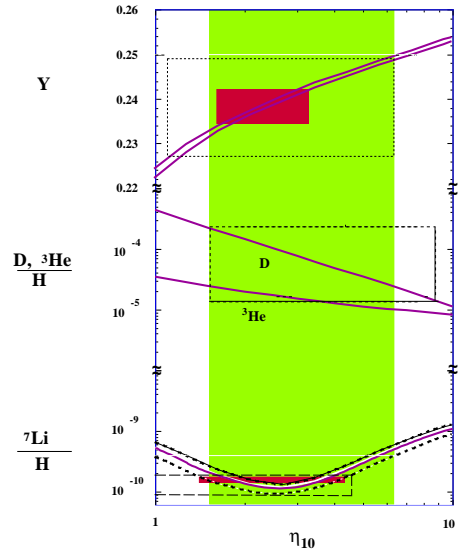}
\caption{
There is good concordance between the observed abundances of light 
elements and calculations of Big-Bang nucleosynthesis~\cite{BBN}.}
\label{fig:BBN}
\end{figure}

The second piece of evidence for the Big Bang was provided by the
abundances of light elements seen in Fig.~\ref{fig:BBN}, which are thought
to have been established when the Universe was about $10^8$ times smaller
and hotter than it is today, with age $\sim 1$ {\rm to} $10^2$~s. This
nuclear `cooking' must have occurred when the temperature $T$ of the
Universe corresponded to characteristic particle energies $\sim 1$~MeV.

\begin{figure}[htb]
\centering
\includegraphics*[width=65mm]{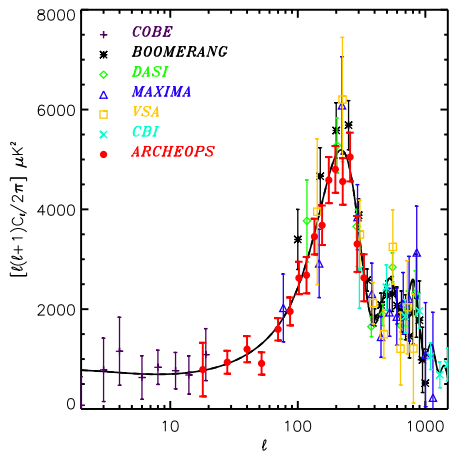}
\caption{
A compilation of data on fluctuations in the cosmic microwave background 
radiation~\cite{ARCHEOPS}.}
\label{fig:CMB}
\end{figure}

Before this time, when it was $\sim 10^{-6}$ to $\sim 10^{-5}$~s old, it
is thought that the Universe made a transition from quarks and gluons to
hadrons at a temperature $T \sim 100$~MeV. Previous to that, the
electroweak transition when Standard Model particles acquired their masses
would have occurred when the Universe was $\sim 10^{-12}$ to $\sim
10^{-10}$~s old, and the temperature $T \sim 100$~GeV. The CMB is thought
to provide a window to an even earlier epoch, via its small-scale
fluctuations $\delta \rho / \rho$, which show up at the $10^{-5}$ level,
as seen in Fig.~\ref{fig:CMB}. If these are due to quantum fluctuations
during an inflationary epoch, $\delta \rho / \rho \sim (T
/m_P)^2$, telling us that the typical energies of particles in the
Universe may once have approached $10^{16}$~GeV.

The fact that the CMB fluctuation spectrum is largest for partial waves
$\ell \sim 200$, as seen in Fig.~\ref{fig:CMB}~\cite{ARCHEOPS}, suggests
that the total energy density of the Universe is close to the critical
density: $\Omega_{tot} \sim 1$.  On the other hand, the Big-Bang
nucleosynthesis calculations shown in Fig.~\ref{fig:BBN} indicate that the
density of baryons in the Universe today is far less than the critical
density:  $\Omega_B \equiv \rho_B/\rho_{crit} \sim {\rm few}$~\%, a range
supported by observations of the smaller-scale peaks in the CMB
fluctuations shown in Fig.~\ref{fig:CMB}.  What form does the missing
energy take?

Observations of large-scale structure suggest that the total matter
density $\Omega_m \sim 0.3$, a value supported by a
combination~\cite{concord} of data on CMB fluctuations and high-redshift
supernovae~\cite{SN}. Most of this $\Omega_m$ is thought to be dark
non-baryonic matter. What is the nature of this dark matter?

It certainly includes neutrinos, which are now thought to have non-zero
masses~\cite{SK,SNO}, but these are probably insufficient to explain most
of the dark matter. In any case, people who model the formation of
large-scale structures would prefer more massive `cold' dark matter
particles that would have been non-relativistic when these structures
began to grow. In addition to this cold dark matter, the CMB and other
data apparently require about 2/3 of the critical density to be in the
form of `dark energy' in the vacuum~\cite{SN,concord}, but I do not
discuss the latter further in this talk.  Instead, I concentrate on the
dark-matter particles that might have signatures among high-energy cosmic
rays.

\section{Astrophysical Neutrinos}

Let us first consider generic features of the density of relic neutrinos
or similar neutral weakly-interacting particles. If neutrinos weigh less
than $\sim 1$~MeV, their cosmological relic number density $n_\nu$ was
fixed when $T \sim 1$~MeV and is essentially independent of their mass,
hence their relic density $\rho_\nu = m_\nu n_\nu$ increases linearly with
mass, rising above the critical density when $m_\nu \sim 30$~eV. The
neutrino mass density would be excessive for masses up to $\sim 3$~GeV,
where a Boltzmann factor suppresses the $\nu$ number density sufficiently
to push $\Omega_\nu$ back below unity. The `neutrino' density would be
most suppressed for $m_\nu \sim m_Z / 2$, when `relic' annihilation is
most efficient. For larger `neutrino' masses, the annihilation rate
typically falls and the relic density (which is fixed when $T \sim m_\nu /
25$) correspondingly rises, reaching the critical density for some
`neutrino' mass $\sim 1$~TeV.

Thus, there are three mass ranges where such a `neutrino' might have a
relic density of interest for astrophysics and cosmology: when $m_\nu \sim
10$~eV, $\sim 3$~GeV or $\sim 100$~GeV to 1~TeV. In the first of these
windows, the neutrino would constitute hot dark matter, in the latter two
it would be cold dark matter. The middle option is excluded by a
combination of experiments at LEP and direct dark-matter searches, and the
third option is that exercised by the lightest supersymmetric
particle~\cite{EHNOS}, as we discuss later.

\begin{figure}[htb]
\centering
\includegraphics*[width=65mm]{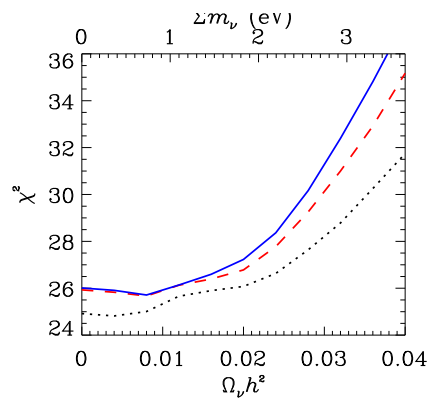}
\caption{
Cosmological upper limits on the neutrino mass~\cite{Hannestad}, based on 
various 
combinations of data on the cosmic microwave background radiation and 
large-scale structures in the Universe.}
\label{fig:neutrino}
\end{figure}

Observations of large-scale structures in the Universe favour the 
predominance of cold dark matter, as already mentioned, and can be used to 
set an upper limit on the sum of light neutrino masses~\cite{Hannestad}:
\begin{equation}
\Sigma_i m_{\nu_i} \; \; < \; \; 3 \; {\rm eV},
\label{cosmonu}
\end{equation}
as seen in Fig.~\ref{fig:neutrino}.
The neutrino oscillation experiments force all three neutrino flavours to 
be essentially degenerate compared to (\ref{cosmonu}), implying that 
$m_{\nu_i} < 1$~eV for each species. This is stronger than the upper 
limit $m_{\nu_e} < 2.5$~eV coming from the end-point of Tritium $\beta$ 
decay~\cite{Tritium}, and future observations of large-scale structure 
should improve the sensitivity to $m_\nu \sim 0.3$~eV. 

\begin{figure}[htb]
\centering
\includegraphics*[width=65mm]{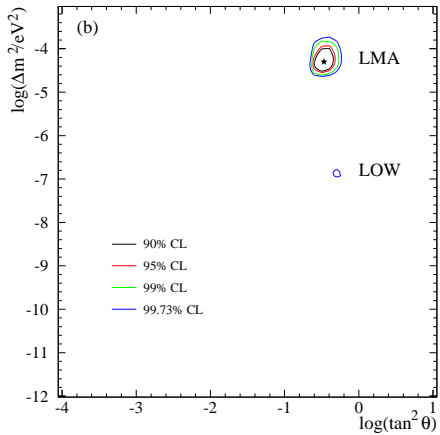}
\caption{
A global fit to solar neutrino data, following the SNO measurements of 
the total neutral-current reaction rate, the energy spectrum and the 
day-night asymmetry, favours large mixing and $\Delta m^2 \sim 6 \times 
10^{-5}$~eV$^2$~\cite{SNO}.} 
\label{fig:SNO}
\end{figure}

Astrophysical sources~\cite{Nobel} have provided us with the first
confirmed evidence for neutrino oscillations and (presumably) neutrino
masses. The first astronomical image obtained with neutrinos was that of
the Sun, via neutrino-electron scattering. Recently, comparative
measurements of charged-current and neutral-current reactions by
SNO~\cite{SNO} have established beyond any doubt that solar $\nu_e$
oscillate into some combination of $\nu_\mu$ and $\nu_\tau$, most probably
with a relatively large mass-squared difference $\Delta m^2 \sim 6 \times
10^{-5}$~eV$^2$ and relatively large, but not maximal, mixing $\sin^2 2
\theta \sim 0.8$ (the LMA solution), as seen in
Fig.~\ref{fig:SNO}~\cite{SNO}.

\begin{figure}[htb]
\centering
\includegraphics*[width=65mm]{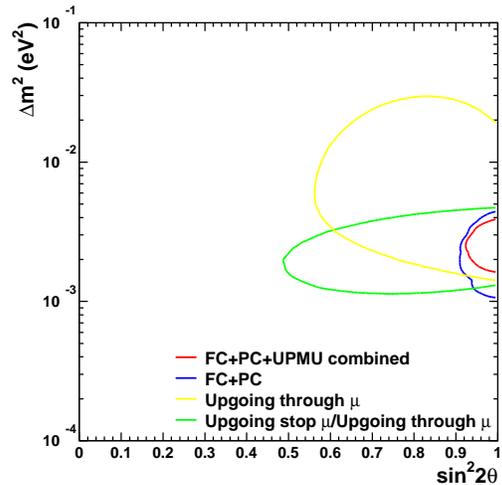}
\caption{A fit to the Super-Kamiokande data on atmospheric 
neutrinos~\cite{SK} indicates near-maximal $\nu_\mu - \nu_\tau$ mixing 
with $\Delta m^2 \sim 2.5 \times 10^{-3}$~eV$^2$.} 
\label{fig:SK}
\end{figure}

As you know, atmospheric neutrinos are produced by cosmic rays with
energies mainly in the range $\sim 1 - 10$~GeV. For some time now, it has
been established that $\nu_\mu$ also oscillate~\cite{SK}, probably mainly
into $\nu_\tau$ with $\Delta m^2 \sim 2.5 \times 10^{-3}$~eV$^2$ and
near-maximal mixing $\sin^2 2 \theta \sim 1$, as seen in
Fig.~\ref{fig:SK}. The available data on the primary cosmic-ray spectrum
and particle production at accelerators enable the atmospheric neutrino
flux to be calculated quite reliably, and further improvements will be
possible using the data presented here from the AMS~\cite{AMS} and
L3+C~\cite{L3+C} Collaborations.

\begin{figure}[htb]
\centering
\includegraphics*[width=65mm]{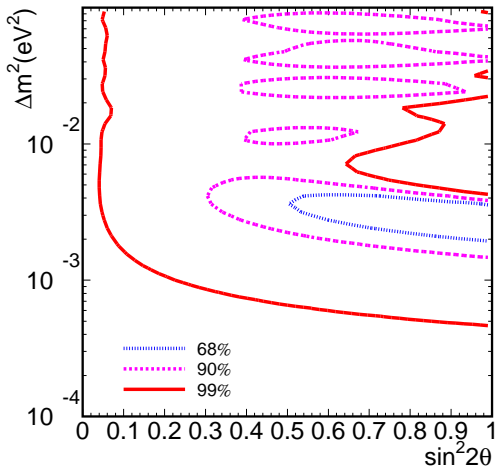}
\caption{
The region of $\Delta m^2$ and the mixing angle favoured by preliminary 
data from the K2K long-baseline experiment is highly consistent 
with the Super-Kamiokande data shown in Fig.~\ref{fig:SK}~\cite{K2K}.}
\label{fig:K2K}
\end{figure}

Long-baseline neutrino oscillation experiments are underway to refine and
extend these astrophysical neutrino oscillation results. The K2K
experiment already provides some confirmation of atmospheric $\nu_\mu$ 
oscillations,
as seen in Fig.~\ref{fig:K2K}~\cite{K2K}, the KamLAND experiment is
expected soon to test definitively the LMA solution for solar
neutrinos~\cite{KamLAND}, the MINOS experiment will look explicitly for
the oscillatory pattern and have improved sensitivity to $\nu_\mu \to
\nu_e$ oscillations~\cite{MINOS}, and the OPERA and ICARUS experiments
should be able to observe the $\tau$ production expected following
$\nu_\mu \to \nu_\tau$ oscillations~\cite{CNGS}.

\section{Supersymmetry}

As already mentioned, cosmic rays in the range from 100~GeV to 1~TeV may
be the best place to look for supersymmetry: why? The primary theoretical
motivation for expecting supersymmetry to appear in this energy range is
provided by the hierarchy problem~\cite{hierarchy}: why is the electroweak
scale $m_W$ so much less than the Planck scale $m_P \sim
10^{19}$~GeV, which is the only candidate we have for a fundamental mass
scale in physics? Equivalently, why is $G_F \sim 1 / m_W^2 \ll G_N = 1 /
m_P^2$?

You might just say, why not choose the value of $m_W$ and forget about the 
problem? Life is not as simple as that, because quantum effects in the 
Standard Model make very large corrections to the electroweak scale:
\begin{equation}
\delta m_W^2 \; \sim \; {\cal O}({\alpha \over \pi}) \Lambda^2,
\label{Lambda}
\end{equation}
where $\Lambda$ is a cutoff representing the scale at which the Standard 
Model must be modified by introducing new physics. A mechanism for 
cutting the divergence (\ref{Lambda}) off in a natural way is provided by 
supersymmetry, which exploits the opposite signs in the 
quadratically divergent fermionic and bosonic corrections to the 
electroweak scale:
\begin{equation}
\delta m_W^2 \; \sim \; {\cal O}({\alpha \over \pi}) | m_B^2 - m_F^2|,
\label{natural}
\end{equation}
which $\sim m_W^2$ if $| m_B^2 - m_F^2| \sim 1$~TeV$^2$. This argument 
therefore leads one to expects supersymmetric partners of Standard Model 
particles to appear at or below the TeV scale~\cite{hierarchy}.

\begin{figure}[htb]
\begin{flushleft}
\includegraphics*[width=150mm]{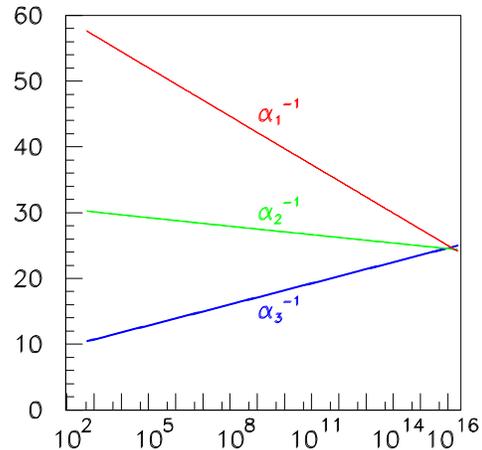}
\end{flushleft}
\vspace{-8cm}
\caption{The measurements (vertical axis) of the gauge coupling strengths
of the
Standard Model at LEP and elsewhere can be evolved up to high energies
(horizontal axis, in units of GeV)
using renormalization-group equations incorporating supersymmetry. They
are consistent with unification at a very high energy scale, but not with
unification without supersymmetry~\cite{GUT}.}
\label{fig:GUT}
\end{figure}

This argument for low-energy supersymmetry is supported circumstantially
by the possibility it offers for unification of the gauge couplings.  
Such grand unification does not occur in the absence of supersymmetry, but
is quite possible if supersymmetric particles weigh about 1~TeV, as seen
in Fig.~\ref{fig:GUT}~\cite{GUT}.

In many supersymmetric models, the lightest supersymmetric particle (LSP)
$\chi$ is stable, and a good candidate for cold dark matter~\cite{EHNOS}, 
which provides
a third general argument for the TeV mass scale. The relic energy density
$\rho_\chi = m_\chi n_\chi$, where the relic number density
\begin{equation}
n_\chi \; \sim \; {1 \over \sigma_{ann} (\chi \chi \to {\rm all})},
\end{equation}
where a typical annihilation cross section $\sigma_{ann} \sim 1 / 
m_\chi^2$. Thus, the overall relic density increases with mass, and 
typically becomes too high when $m_\chi > 1$~TeV.

\begin{figure}[htb]
\centering
\includegraphics*[width=65mm]{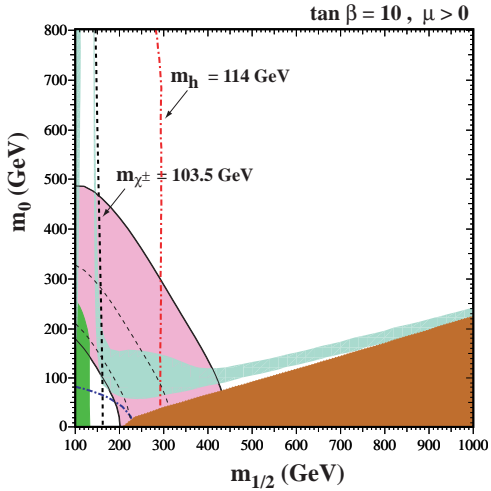}
\caption{The parameter space of the MSSM projected onto the $(m_{1/2}, 
m_0)$
plane for $\tan \beta = 10$ and $\mu > 0$. The LEP lower limits on the
Higgs, chargino and selectron masses are shown as (red) dot-dashed,
(black) dashed and (blue) dash-dotted lines, respectively. The region
at small $(m_{1/2}, m_0)$ excluded by $b \to s \gamma$ is shaded
(green). The dark (red) shaded
region is excluded because dark matter must be neutral, and the region
where its relic density falls within the range preferred by cosmology has 
light (turquoise) shading. The region preferred by the BNL measurement of
$g_\mu -2$ and low-energy $e^+e^-$ data is shaded (pink)~\cite{EFOS}.} 
\label{fig:CMSSM}
\end{figure}

However, the dark-matter annihilation rate may in exceptional
circumstances be enhanced, reducing the supersymmetric relic density for a
given mass, and thereby allowing larger relic masses. For example, if the
LSP and the next-to-lightest supersymmetric particle ${\tilde X}$ have
similar masses, $n_\chi$ may be suppressed by coannihilation
processes~\cite{EFO}: $\sigma (\chi {\tilde X} \to {\rm all})$, which can
be important if $m_{\tilde X} - m_\chi / m_\chi \sim 1 / 10$. This
coannihilation mechanism can provide an allowed `tail' of parameter space
extending out to larger $m_\chi$, as seen in Fig.~\ref{fig:CMSSM}. Such a
tail may also happen when rapid annihilation through a direct-channel pole
is possible, for example if $m_\chi \sim m_{H, Z, ...} /2$~\cite{pole}.

The space of input supersymmetric fermion masses $m_{1/2}$ and boson
masses $m_0$ is illustrated for one particular ratio $\tan \beta$ of the
Higgs vacuum expectation values in the minimal supersymmetric extension of
the Standard Model (MSSM) in Fig.~\ref{fig:CMSSM}~\cite{EFOS}. We see at 
large
values of the ratio $m_{1/2} / m_0$ a region excluded because there the
relic particle would be charged, a possibility excluded by astrophysics.
At small $m_{1/2}$ and/or $m_0$ we see experimental exclusions from the
absences of the supersymmetric partners of the electron $\tilde e$ and of
the $W/H^\pm$, and also of the Higgs boson $H$. A dark (green) shaded 
region is
excluded by measurements of $b \to s \gamma$ decay, and a lighter 
(pink) shading
shows regions favoured by the recent measurement of the anomalous magnetic
moment of the muon~\cite{BNL}.

Finally, the lightest (turquoise) shading in Fig.~\ref{fig:CMSSM} picks 
out the region
where the relic LSP density lies within the range favoured by cosmology:  
$0.1 < \Omega_\chi h^2 < 0.3$. We see that much of this region is
disfavoured by the accelerator constraints, particularly the LEP Higgs
limit $m_H > 114$~GeV~\cite{LEPHWG}. 

\begin{figure}[htb]
\centering
\includegraphics*[width=65mm]{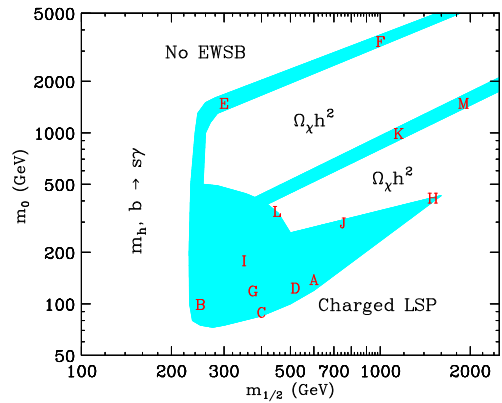}
\caption{Sketch of the distribution of proposed CMSSM benchmark points 
in the $(m_{1/2}, m_0)$ plane~\cite{Bench}. These points were chosen as
illustrations of the range of possibilities in the CMSSM, rather than as
a `fair' sample of its parameter space.}
\label{fig:bench}
\end{figure}

A set of `benchmark' supersymmetric scenarios was recently
proposed~\cite{Bench}, that respect all the experimental and cosmological
constraints on the minimal supersymmetric extension of the Standard Model
(MSSM). As shown in Fig.~\ref{fig:bench}, they indicate the range of
options, rather than sample the parameter space in a `fair' manner. The
LHC has great possibilities for detecting supersymmetry, principally via
events with missing energy and other signatures such as high-energy
leptons and/or jets. These prospects for discovering supersymmetry may be
compared with those for detecting astrophysical supersmmetric dark matter.

\section{Search for Supersymmetric Dark Matter}

Searches among cosmic rays with energies up to about 1~TeV provide several
promising signatures for supersymmetric dark matter particles, that may
enable this community to `scoop' the LHC~\cite{DMreview}.

\begin{figure}[htb]
\centering
\includegraphics*[width=65mm]{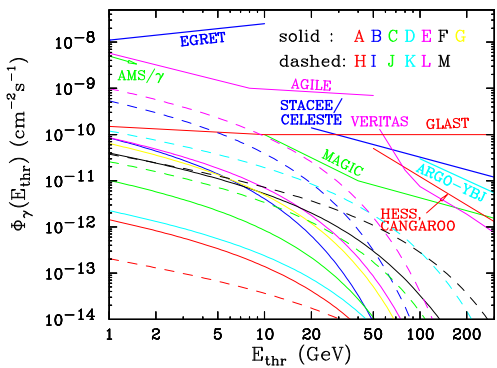}
\caption{
Observations of $\gamma$ rays from the galactic centre by GLAST and 
ground-based experiments may be able to test certain 
supersymmetric benchmark scenarios~\cite{EFFMO}.}
\label{fig:gamma}
\end{figure}

One possibility is to look for energetic {\it gamma rays} that may be
emitted by LSP annihilations in the core of the Milky Way. The benchmark
models indicate that these might have typical energies $\sim 10$~GeV, and
detectors such as GLAST with a threshold as low as $\sim 1$~GeV might have
a better chance, as seen in Fig.~\ref{fig:gamma}~\cite{EFFMO}. The
prospects for these searches depend on the degree to which the dark-matter
particle density may be enhanced in the core of the Milky Way, which is
uncertain by orders of magnitude. In Fig.~\ref{fig:gamma}, a
middle-of-the-road enhancement by a factor 200 has been assumed.

\begin{figure}[htb]
\centering
\includegraphics*[width=65mm]{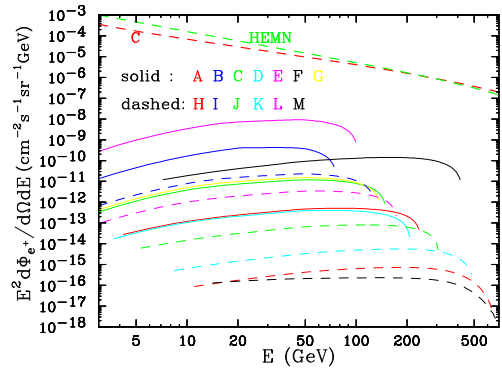}
\caption{
Comparison between the cosmic-ray positron background and the fluxes
from the annihilations of relic particles, calculated~\cite{EFFMO} in
supersymmetric benchmark scenarios.}
\label{fig:positron}
\end{figure}

Another possibility is to look for {\it positrons} emitted by LSP
annihilations in the halo of the Milky Way. In this case, the benchmark
models indicate that energies $\sim 100$~GeV might be the most
interesting, though the signal may be less promising than in the $\gamma$
case, as seen in Fig.~\ref{fig:positron}~\cite{EFFMO}.

\begin{figure}[htb]
\centering
\includegraphics*[width=65mm]{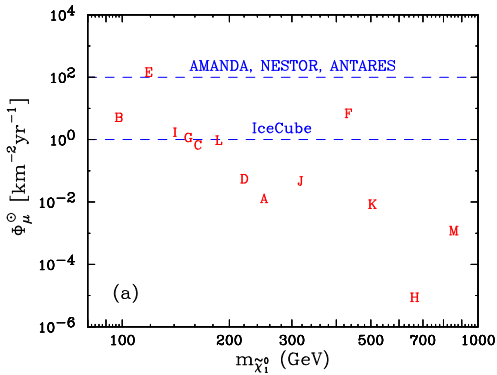}
\caption{
Searches in IceCube and other km$^2$ detectors for energetic muons 
originating from the interactions of high-energy 
neutrinos produced by the annihilations of supersymmetric relic particles 
captured inside the Sun may probe some supersymmetric benchmark 
scenarios~\cite{AMANDA}.}
\label{fig:muon}
\end{figure}

One of the most promising signatures is {\it energetic muons} produced in
the Earth by energetic neutrinos emitted by LSP annihilations in the
centre of the Sun or Earth. In this case, as seen in Fig.~\ref{fig:muon},
all energies up to $\sim 1$~TeV might be important. According to our
calculations~\cite{EFFMO}, the prospects for detecting relic annihilations
in the core of the Sun appear more promising than those in the centre of
the Earth. Some upper limits on the energetic solar muon flux have already
been produced, most recently by the AMANDA Collaboration~\cite{AMANDA},
which already begin to exclude some more extreme supersymmetric models.

\begin{figure}[htb]
\centering
\includegraphics*[width=65mm]{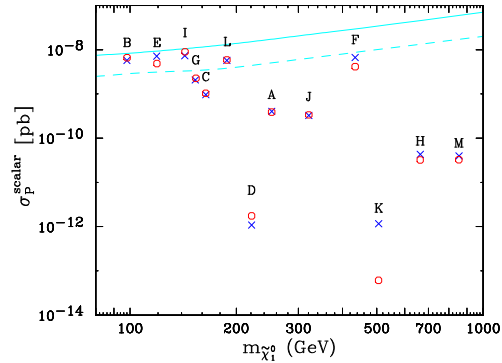}
\caption{
Direct searches for the scattering of superysmmetric relic 
particles in underground detectors may probe some supersymmetric benchmark
scenarios~\cite{EFFMO}.}
\label{fig:direct}
\end{figure}

The most convincing evidence for supersymmetric dark matter might
eventually come from direct searches for the {\it scattering of relic
particles} on nuclei in the laboratory~\cite{GoodWitt}. Here the best
chances seem to be offered by spin-independent scattering on relatively
heavy nuclei. There has been a claim by the DAMA
Collaboration~\cite{DAMA} to have observed an annual modulation effect due
to the scattering of dark matter particles, but this interpretation of
their data has been largely excluded by the CDMS~\cite{CDMS},
EDELWEISS~\cite{EDELWEISS} and UKDMC experiments. In any
case, reproducing the DAMA data would have required a scattering cross
section much larger than predicted in the simple supersymmetric models
studied in~\cite{EFFMO} and~\cite{EFerstlO}. As illustrated in
Fig.~\ref{fig:direct}, future large cryogenic detectors, such as that
proposed by the Heidelberg group, would good chances in many
supersymmetric scenarios.

\section{Space-Time Foam}

We know that space-time is essentially flat at large scales, but 
quantum gravitational 
effects are expected to cause large fluctuations on small scales in length 
and time:
\begin{equation}
\Delta E, \Delta \chi = {\cal O}(1) \; \; {\rm in} \; \; \Delta x, \Delta 
t = {\cal O}(1),
\label{flux}
\end{equation}
where the energy $E$, distance $x$ and time $t$ are all measured in Planck 
units $\sim 10^{19}$~GeV,  $\sim 10^{-33}$~cm and  $\sim 10^{-43}$~s, 
respectively, and $\chi$ is a generic dimensionless measure of topology. 
Are there any observable consequences of such quantum gravitational 
effects?

One suggestion has been that information might be lost across microscopic 
event horizons associated with such topological fluctuations, causing an 
apparent modification of microscopic quantum 
mechanics~\cite{Hawking,EHNS,EMN}. Another has been 
that the gravitational recoil of the vacuum as an energetic particle 
passes by might modify special relativity, in such a way that the 
particle's velocity might be reduced~\cite{AEMN,AEMNS}:
\begin{equation}
c(E) \; = \; c \times \left( 1 - {E \over M} + \cdots \right),
\label{Lorentz}
\end{equation}
where $M$ is a quantum-gravity mass scale that might $\sim m_P \sim 
10^{19}$~GeV.

The best probes of the possibility (\ref{Lorentz}) may be provided by 
astrophysical sources of $\gamma$ rays: they have large distances $D$ and 
hence light propagation times $t = D / c$, and hence their light pulses 
may exhibit time delays
\begin{equation}
\Delta t \; \sim \; - {D \over c^2} (c(E) - c).
\end{equation}
Distant high-energy sources with short characteristic time scales 
$\delta t$ have the best sensitivity to the quantum-gravity 
scale~\cite{AEMNS}:
\begin{equation}
M \; \sim \; {E \cdot D \over \delta t}.
\end{equation}
{}From this point of view, interesting astrophysical sources include 
pulsars (relatively short distances, but short time scales), AGNs 
(moderate redshifts $z$ and relatively large time 
scales, but high energies) and GRBs ($z \sim 1$, 
$\delta t \sim 10^{-2}$~s). There are confirmed observations of GRBs only 
at relatively low energies, but there have been reports of 
GeV or even TeV~\cite{LANL} photons emitted by GRBs.

\begin{figure}[htb]
\centering
\includegraphics*[width=65mm]{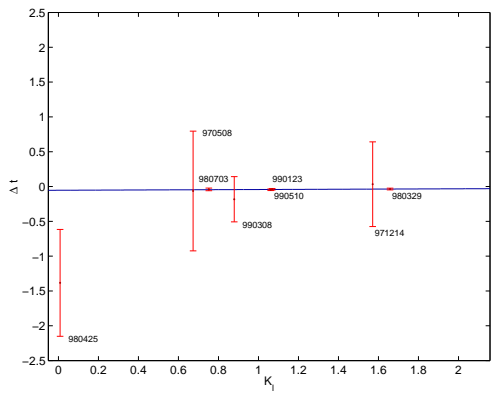}
\caption{
Search for a correlation with distance $\propto K_l$ of the time-lags 
$\Delta t$ between structures in GRB
emissions at different energies, as observed by the BATSE and OSSE 
instruments on the CGRO~\cite{EMNS}.}
\label{fig:EMNS}
\end{figure}

Some time ago, we published an analysis of all the GRBs whose redshifts
had then been measured, using data from the BATSE and OSSE instruments on
the Compton Gamma-Ray Observatory (CGRO), finding no significant
correlation of time-lag with $z$, and inferring that $M \gappeq
10^{15}$~GeV~\cite{EFMMN}. More recently, we have been making an improved
analysis, benefiting from the measurements of more GRB redshifts, and
using TTE data from BATSE, which has a finer time resolution, and making a
wavelet analysis, whose result is shown in Fig.~\ref{fig:EMNS}.  
This time, we find $M \gappeq 7.9 \times 10^{15}$~GeV~\cite{EMNS}.
In the future, GLAST should be able to improve significantly on this
sensitivity~\cite{GLAST}.

\section{High-Energy Cosmic Rays and the LHC}

Colliding pairs of protons with 14~TeV each, the LHC will be equivalent to
a beam of particles with energies $10^{17}$~GeV striking a fixed target.  
The general-purpose detectors, ATLAS~\cite{ATLAS} and CMS~\cite{CMS}, are
designed to see `everything' emitted with a centre-of-mass rapidity $|
\eta | \lappeq 3$.  The LHCb experiment~\cite{LHCb} will also be able to
measure some diffractive physics. The LHC will also be able to collide
beams of heavy ions with energies $\sim 5$~TeV per nucleon, with the ALICE
experiment~\cite{ALICE} as principal user, and could in principle also
collide protons or deuterons with heavy ions. Also planned is the TOTEM
experiment~\cite{TOTEM} to measure the total and elastic $pp$ cross
sections, as well as some diffractive inelastic events. This experiment
will have some capabilities to measure particles with $| \eta | \lappeq 3$
or 5, and the LHC physics community would like to hear from the cosmic-ray
community what its needs are~\cite{NEEDS}. This is the subject of a
special workshop at this meeting: we hope that the LHC can contribute to
refining models of ultra-high-energy cosmic rays. We should like to know
what particles the cosmic-ray community would like to see measured in what
rapidity range.

\section{Ultra-High-Energy Cosmic Rays}

\begin{figure}[htb]
\centering
\includegraphics*[width=65mm]{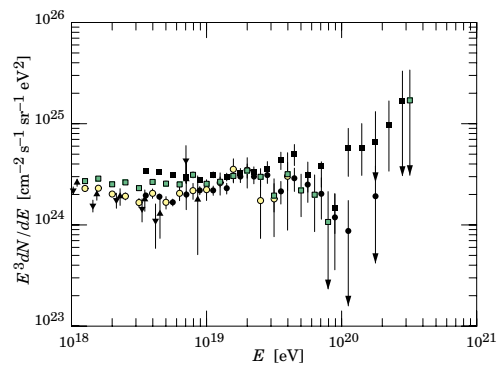}
\caption{
Compilation of data on ultra-high-energy cosmic rays~\cite{Tritium}. AGASA 
and other experiments have reported some events beyond the GZK 
cutoff~\cite{AGASA}, but 
the HiRes experiment~\cite{HiRes} (lighter symbols) reports no significant 
excess.}
\label{fig:UHECR}
\end{figure}

As you know, extended-air-shower (EAS) experiments~\cite{AGASA} have
reported an apparent excess of events beyond the GZK cutoff, but this is
not confirmed by HiRes~\cite{HiRes}, the largest fluorescence detector, as
seen in Fig.~\ref{fig:UHECR}. At issue are the absolute and relative
energy calibrations of the EAS and fluorescence techniques. The former
depends on the models that we hope the LHC will help refine, and the
latter rely on normalizations of fluoresecnce lines around 390~nm.  A
recent experiment seems to find a discrepancy with previous measurements
in this region, and the apparent conflict should be
clarified~\cite{Watson}.

There are two main classes of models for ultra-high-energy cosmic rays 
(UHECR): `bottom-up' and `top-down'. The former postulate acceleration by 
astrophysical sources, which should have a minimum size $R$:
\begin{equation}
R \; \sim \; 
{100 \over Z} 
\left( 
{E \over 10^{20}~{\rm eV} } \right) 
\left( { \mu{\rm G} \over B} \right),
\label{Hillas}
\end{equation}
where $Z$ is the atomic number, $E$ is the 
energy and $B$ is the magnetic field bending the UHECR. Alternatively, the 
maximum energy attainable in any given source is
\begin{equation}
E \; \sim \; 10^{18} Z \left( { R \over {\rm Kpc}} \right) \left( {B \over
\mu{\rm G}} \right) {\rm eV}.
\end{equation}
Popular candidate sources include GRBs and AGNs, with neutron 
stars and colliding galaxies also proposed.

The alternative class of `top-down' models postulates the production of
UHECR by GUT-scale physics, such as topological defects or the decays of
metastable superheavy relic particles~\cite{cryptons} - which have some
chances of being produced with interesting cosmological relic
densities~\cite{Kolb}. These models predict some anisotropy if the
particles have a typical halo distribution~\cite{Sarkar}, and could also
exhibit clustering if the halo is lumpy~\cite{Blasi}. A characteristic of
`top-down' models is that they predict large numbers of photons among the
UHECR, and possibly even supersymmetric particles~\cite{UHECRsusy}!

Other suggestions for UHECR include the collisions of ultra-high-energy
neutrinos with non-relativistic massive neutrinos to produce $Z$
bosons~\cite{Zburst,Fodor}.  It has also been suggested~\cite{GM} that the
type of modification of Lorentz kinematics mentioned earlier might also
help distant sources evade the GZK cutoff.

There will be much discussion of these ideas during this meeting. We are
all glad that the Auger experiment in Argentina is progressing
well~\cite{Dova}. Its high statistics and combination of the EAS and
fluorescence techniques should pin down the existence of UHECR and
discriminate between rival models. In the longer run, the EUSO
project~\cite{EUSO} would be able to provide even higher statistics.

\section{Conclusions}

This talk has provided only a brief review of the interfaces between
high-energy physics, astrophysics and cosmology. As we have seen,
high-energy cosmic-ray physics may provide much useful information to
particle physicists, for example concerning neutrinos and dark matter, and
possibly even quantum gravity. Likewise, accelerator experiments can
provide useful input for cosmic-ray experiments, for example by testing,
calibrating and validating simulation codes.

During this talk, we have discussed both localized astrophysical sources
of high-energy particles and global sources provided by the very early
Universe. Although lower-energy cosmic rays are believed to originate from
local sources within our own galaxy, the origin of the highest-energy
cosmic rays is still unknown. Both astrophysical sources and cosmological
origins are being actively proposed. Unravelling the origins of the UHECR
will surely require an active dialogue between accelerator and cosmic-ray
physicists.

\end{document}